\documentclass[fleqn,usenatbib,useAMS]{mnras}

\usepackage{newtxtext,newtxmath}

\usepackage[T1]{fontenc}

\DeclareRobustCommand{\VAN}[3]{#2}
\let\VANthebibliography\thebibliography
\def\thebibliography{\DeclareRobustCommand{\VAN}[3]{##3}\VANthebibliography}

\usepackage{graphicx}	
\usepackage{amsmath}	

\defcitealias{PaperI}{Paper I}

\title[Black widow formation]{Black widow formation by pulsar irradiation and sustained magnetic braking}

\author[S. Ginzburg and E. Quataert]{
Sivan Ginzburg$^{1}$\thanks{E-mail: ginzburg@berkeley.edu}\thanks{51 Pegasi b Fellow.}
and Eliot Quataert$^{1,2}$
\\
$^{1}$Department of Astronomy and Theoretical Astrophysics Center, University of California, Berkeley, CA 94720, USA\\
$^{2}$Department of Astrophysical Sciences, Princeton University, Princeton, NJ 08544, USA
}

\date{Accepted XXX. Received YYY; in original form ZZZ}

\pubyear{2020}

\begin{document}
\label{firstpage}
\pagerange{\pageref{firstpage}--\pageref{lastpage}}
\maketitle

\begin{abstract}
Black widows are millisecond pulsars with low-mass companions, a few per cent the mass of the sun, on orbits of several hours. These companions are presumably the remnants of main sequence stars that lost their mass through a combination of Roche-lobe overflow and ablation by the host pulsar's high-energy radiation. While ablation itself is too weak to significantly reduce the mass of the companion star, the ablated wind couples to its magnetic field, removes orbital angular momentum, and thus maintains stable Roche-lobe overflow. We use the \textsc{mesa} stellar evolution code, complemented by analytic estimates, to track initially main sequence companions as they are reduced to a fraction of their original mass by this ablation-driven magnetic braking. 
We argue that magnetic braking remains effective even for low-mass companions.
A key ingredient of our model is that the irradiating luminosity of the pulsar $L_{\rm irr}$ deposits energy in the companion's atmosphere and thereby slows down its Kelvin--Helmholtz cooling. We find that the high-energy luminosities measured by {\it Fermi} $L_{\rm irr}=0.1-3\,{\rm L_{\sun}}$ can explain the span of black widow orbital periods. The same $L_{\rm irr}$ range reproduces the companions' night-side temperatures, which cluster around 3000 K, as inferred from optical light curves. 
\end{abstract}




\section{Introduction}\label{sec:intro}

Over the last decade, the number of eclipsing millisecond radio pulsars has greatly increased, thanks to dedicated surveys and followup of {\it Fermi} $\gamma$-ray sources \citep{Keith2010, Bates2011, Ray2012, Roberts2013}. These systems are often divided into `black widows' --- with companions a few per cent the mass of the sun (${\rm M}_{\sun}$), and `redbacks' --- with companions more massive than about $0.1 {\rm M}_{\sun}$ \citep{Chen2013,Roberts2013,DeVito2020}.

The companions of black widows and redbacks (collectively known as `spiders') are believed to be the remnants of main sequence stars that lost orbital angular momentum and were driven to Roche-lobe overflow by magnetic braking --- spinning up their host pulsars to millisecond periods in the process \citep{Phinney88, Benvenuto2012, Chen2013}. It is usually assumed that magnetic braking halts when the companion becomes fully convective, and from then onward it loses mass primarily through ablation by the pulsar's high-energy irradiation \citep{Chen2013, DeVito2020}. The ablated wind, which is ultimately powered by the pulsar's spin down, may explain the observed eclipses in these systems \citep{Kluzniak1988,Phinney88}.

Despite its pivotal role in black widow evolution, the evaporative wind is conventionally modelled with a simple linear relation, linking the mass loss rate to the pulsar's spin-down power using an unknown efficiency parameter \citep{Benvenuto2012,Benvenuto2014,Benvenuto2015MNRAS,Chen2013,JiaLi2015,JiaLi2016,LiuLi2017,Ablimit2019}. In \citet[][hereafter \citetalias{PaperI}]{PaperI}, on the other hand, we calculated the companion's mass loss rate explicitly by studying the hydrodynamical \citet{Parker1958} wind launched off its atmosphere. Specifically, we adapted the \citet{Begelman83} analysis of Compton heated winds from accretion discs, and improved upon the estimate of \citet{Ruderman89E} by determining the wind's sonic point more accurately. 
We found that the evaporation efficiency is not constant and is also much lower than typically assumed in recent literature. With the possible exception of a few systems
(which include the original black widow discovered by \citealt{Fruchter1988}, cf. \citealt{EichlerLevinson88,LevinsonEichler91}), evaporation by $\gamma$-rays is on its own too weak to transform stars into black widow companions over a Hubble time \citepalias[eq. 17 in][]{PaperI}. Instead, we suggested in \citetalias{PaperI} that the wind couples to the companion's magnetic field, removes angular momentum from the orbit, and maintains stable Roche-lobe overflow --- significantly amplifying the mass loss rate.

Here, we use the stellar evolution code \textsc{mesa} \citep{Paxton2011,Paxton2013,Paxton2015,Paxton2018,Paxton2019} to study this irradiation-driven magnetic braking mechanism; we argue that it operates down to low companion masses. We compute evolutionary tracks that follow an initially solar mass main sequence companion as it loses mass, eventually reproducing black widow systems. We self-consistently couple the companion's magnetic field to its convective luminosity \citep{Christensen2009} and check when its interior is fully convective. A key ingredient in our calculation is the suppression of the companion's cooling by the optically deep radiative layer induced by the pulsar's incident flux. This effect has been found to play an important role in the evolution of hot Jupiters \citep{Guillot1996,ArrasBildsten2006}, and related effects were also considered in the context of spider pulsars \citep{BildstenChakrabarty2001,Benvenuto2012,Benvenuto2014, DeVito2020}.

The remainder of this paper is organized as follows. In Section \ref{sec:general} we discuss general considerations when reproducing the observed population of spiders, and in Section \ref{sec:model} we present the specifics of our computational model. We examine the resulting evolutionary tracks in Section \ref{sec:results} and then present analytic estimates to complement our \textsc{mesa} results (Section \ref{sec:analytical}). We summarize our main conclusions in Section \ref{sec:conclusions}. 

\section{General considerations}\label{sec:general}

According to \citetalias{PaperI}, evaporation is too weak to remove significant mass from black widow companions. The only other way for the companion to lose mass is to overfill its Roche lobe. We therefore assume that companions are driven towards Roche-lobe overflow which is thence maintained by loss of orbital angular momentum. In this section we discuss general considerations that apply to any angular momentum sink, with specific mechanisms presented in Section \ref{sec:model}. In some binary evolution calculations, the companion does not always strictly fill its Roche lobe, but rather undergoes cyclic mass transfer episodes while most of the time it is detached. Our analysis below is insensitive to this detail, as long as the companion under-fills its Roche lobe only slightly \citep[e.g. fig. 3 of][]{Benvenuto2012}. The assumption that most black widow companions fill or almost fill their Roche lobes is supported by modelling of their optical light curves \citep{Draghis2019}. 

We denote the pulsar's mass with $M$, and the companion's mass, radius, and separation from the pulsar with $m$, $r$, and $a$ respectively. A Roche-lobe filling companion satisfies
\begin{equation}\label{eq:rlobe}
\frac{r}{a}\simeq 0.49\left(\frac{m}{M}\right)^{1/3},
\end{equation}
where for simplicity and consistency with \citetalias{PaperI}, we take the $m\ll M$ limit of \citet{Eggleton83}, which is accurate to within 20 per cent even for our almost equal mass initial main sequence starting point.
For a Keplerian orbit, the orbital period $P$ is a function of the companion's mean density (approximating again $m\ll M$, which is accurate to within 15 per cent even for our heaviest redback companions)
\begin{equation}\label{eq:period_r}
P=2\upi\left(\frac{r^3}{0.49^3Gm}\right)^{1/2},    
\end{equation}
where $G$ is the gravitational constant. Under the assumption of Roche-lobe overflow, we may therefore evolve the companion as a single star in \textsc{mesa}, and use equation \eqref{eq:period_r} to interpret the orbital period from its radius. The inaccuracies introduced by using the $m\ll M$ approximation are acceptable, given the similar uncertainties in the measured values of both $m$ and $M$.

\begin{figure}
\includegraphics[width=\columnwidth]{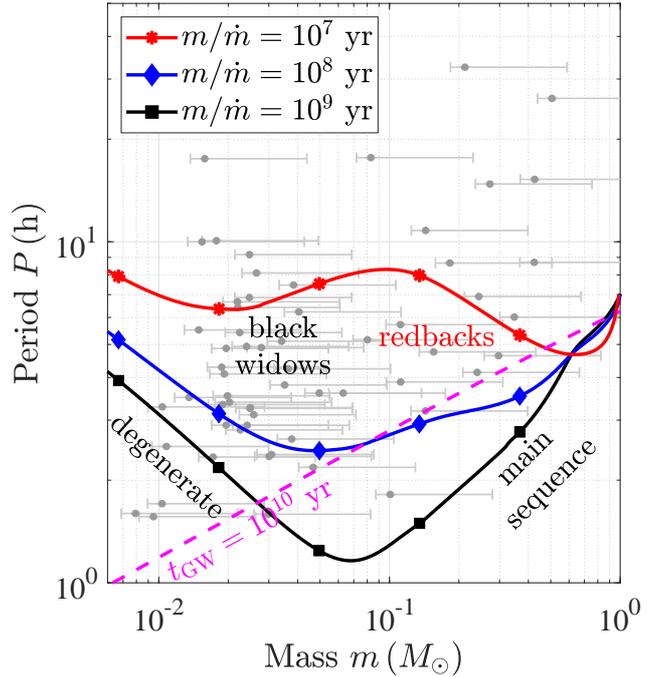}
\caption{Observed spider pulsar population. The grey dots indicate the median companion mass $m$ (inclination angle $i=60^\circ$) with error bars marking the minimum ($i=90^\circ$) and 95 per cent probability ($i=18.2^\circ$) values. The solid curves follow the orbital period evolution of an initially 1 Gyr old, $1 {\rm M}_{\sun}$, Roche-lobe filling companion as it loses mass at different arbitrary rates. The markers indicate each mass e-folding ($10^7$ yr for asterisks, $10^8$ yr for diamonds, and $10^9$ yr for squares). 
The angular momentum loss time due to gravitational waves $t_{\rm GW}$ is indicated for reference (dashed magenta line).
Slowly evaporating companions (black line marked with squares) follow the main sequence before becoming degenerate. Faster mass loss keeps the companions inflated, forming redbacks and black widows. 
The companion's surface boundary condition in this calculation is the same as for an isolated star (i.e. the host pulsar's irradiation is ignored). Under these circumstances, the systems remain in the redback and black widow states only for a short time $10^7-10^8$ yr. Incident irradiation slows down the companion's cooling and is therefore imperative to forming longer-lived spiders (Fig. \ref{fig:tracks}).}
\label{fig:mdot}
\end{figure}

In Fig. \ref{fig:mdot} we present the observed spider pulsar population, taken from the ATNF Pulsar Catalogue  \url{http://www.atnf.csiro.au/research/pulsar/psrcat} \citep{Manchester2005}, version 1.63 (April 2020). Our sample includes all systems with pulsar spin-down periods shorter than 30 ms and non white dwarf companions on orbits shorter than 2 days. This population is often divided into black widows ($m\lesssim 5\times 10^{-2} {\rm M}_{\sun}$) and redbacks ($m\gtrsim 10^{-1} {\rm M}_{\sun}$), with a possible deficit of intermediate mass companions \citep{Roberts2013,Chen2013}. As seen in Fig. \ref{fig:mdot}, the large error bars may reduce the statistical significance of this mass gap. Two pulsars with extremely low mass companions $m\simeq 10^{-3}{\rm M}_{\sun}$: PSR J1719$-$1438 \citep{Bailes2011} and PSR J2322$-$2650 \citep{Spiewak2018}, are not shown in Fig. \ref{fig:mdot} and we consider them outliers.

\begin{figure}
\includegraphics[width=\columnwidth]{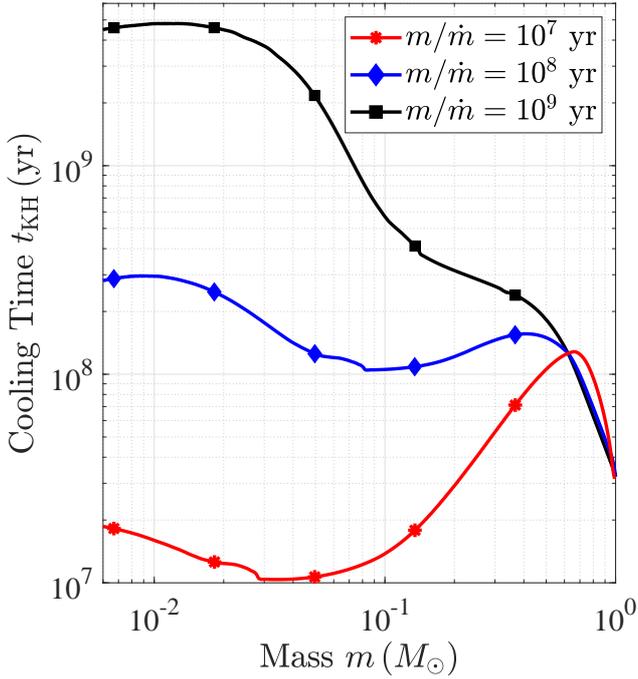}
\caption{The companion's Kelvin--Helmholtz cooling time $t_{\rm KH}$ for the evolutionary tracks presented in Fig. \ref{fig:mdot}. Explicitly, $t_{\rm KH}$ is given by the ratio of the companion's thermal energy to its luminosity. Fast mass loss keeps companions inflated, such that they satisfy $t_{\rm KH}\sim m/\dot{m}$. The life-times of redback and black widow companions that cool down in isolation are short $t_{\rm KH}\ll\textrm{Gyr}$ (red and blue lines, marked with asterisks and diamonds), which is inconsistent with observations. The pulsar's irradiation (not accounted for in this plot) alters the companion's outer boundary condition and thereby lengthens $t_{\rm KH}$, explaining the observed population of longer-lived systems (Figs \ref{fig:tracks} and \ref{fig:timescales}).}
\label{fig:tcool}
\end{figure}

We overlay the observations in Fig. \ref{fig:mdot} with evolutionary tracks of an initially solar mass main sequence star that comes into Roche-lobe contact after 1 Gyr of isolated evolution. Mass is then removed at a variable rate $\dot{m}$, such that the prescribed (and at this point of the paper, arbitrary) mass-loss time-scale $m/\dot{m}$ remains constant.\footnote{Throughout the paper, $\dot{m}$ is to be understood as $|\dot{m}|$, where we omit the absolute value sign for brevity.} Other than that, in this section, the companion evolves as if it were isolated. Specifically, $\dot{m}$ is assumed to already include mass transfer due to all forms of angular momentum loss (gravitational waves, magnetic braking, etc.), and the effects of irradiation on the companion's atmosphere are ignored (these are introduced in Section \ref{sec:irradiation}).
 
The companion's evolution is essentially determined by comparing $m/\dot{m}$ to its Kelvin--Helmholtz cooling time $t_{\rm KH}$, which we plot in Fig. \ref{fig:tcool}. In an intuitive zero-dimensional approximation, the star's pressure $p\sim Gm^2r^{-4}\propto  m^{2/3}\rho^{4/3}$, where $\rho\equiv 3m/(4\upi r^3)$ is the mean density. Adiabatic $(p\propto\rho^\gamma)$ mass loss therefore lowers the density $\rho\propto m^{2/(3\gamma-4)}$, assuming an adiabatic index $\gamma>4/3$, inflating the star beyond its main sequence radius. According to equation \eqref{eq:period_r}, the orbital period increases as $P\propto\rho^{-1/2}\propto m^{-1/(3\gamma-4)}$. However, the mass loss is not adiabatic if the star can radiate away energy and contract. If $t_{\rm KH}<m/\dot{m}$, the companion contracts all the way to the main sequence $P\propto\rho^{-1/2}\propto m^{0.7}$ \citep[assuming approximately $r\propto m^{0.8}$; see][]{Kippenhahn2012}. This is the track followed by the slowly evaporating model (black line, marked with squares) in Fig. \ref{fig:mdot}, which remains close to the main sequence down to the hydrogen burning limit ($\simeq 7\times 10^{-2} {\rm M}_{\sun}$), which is when electron degeneracy becomes important. Further mass loss lowers the density of the nearly degenerate companion, increasing its period according to equation \eqref{eq:period_r}. Companions that lose mass faster do not have enough time to contract onto the main sequence or degeneracy curves, and they remain inflated, with a larger radius (and therefore period) that satisfies $t_{\rm KH}\sim m/\dot{m}$ (red and blue lines, marked with asterisks and diamonds in Figs \ref{fig:mdot} and \ref{fig:tcool}). 

While the high $\dot{m}$ curves in Fig. \ref{fig:mdot} reproduce observed black widows and redbacks, the simulated systems are short lived --- the tracks traverse the spider region in about $10^7-10^8$ yr. This result is at odds with the observed population: if spiders were a short lived phase, they should have been vastly outnumbered by both their main sequence progenitors (this is not the case, as seen in Fig. \ref{fig:mdot}) and by isolated millisecond pulsars, which have completely evaporated their companions \citep[but the numbers are comparable; e.g.][who find the same number of isolated millisecond pulsars and spider pulsars among the {\it Fermi} $\gamma$-ray sources]{Abdo2013}. 
We emphasize that the short spider evolution times do not depend on the specifics of our mass-loss scheme. As Fig. \ref{fig:tcool} shows, the spider companions have short Kelvin--Helmholtz cooling times $t_{\rm KH}\ll\textrm{ Gyr}$. Since $t_{\rm KH}$ is largely a function of $m$ and $r$ (so through equation \ref{eq:period_r}, $t_{\rm KH}$ is a function of $m$ and $P\propto r^{3/2}$), the companions either contract to lower periods on this time-scale (if $m/\dot{m}>t_{\rm KH}$) or lose their mass on an even shorter time (if $m/\dot{m}<t_{\rm KH}$), for any given mass-loss rate $\dot{m}$, regardless of their previous history. A possible solution is that the observed companions have detached from their Roche lobes and contracted to a much smaller radius ($t_{\rm KH}$ lengthens with decreasing radius); this scenario is disfavoured observationally \citep{Draghis2019}.

The tension posed by the short cooling times of black widow and redback companions is naturally resolved when the influence of the host pulsar's irradiation on the companion's atmosphere is taken into account. Typical spider pulsar spin-down luminosities are several times the solar luminosity ${\rm L}_{\sun}$. A significant fraction of this spin-down power is carried by photons \citep{Abdo2013} that can deposit their energy deep in the atmosphere of the short-period companion. The incoming flux changes the atmosphere's outer boundary condition: it induces a deep radiative layer in the otherwise convective envelope of the companion, slowing down its cooling. This effect has been invoked to explain the inflated radii of hot Jupiters with orbital periods of a few days \citep{Guillot1996,ArrasBildsten2006}. As we show below, typical spider companions, with periods of several hours (Fig. \ref{fig:mdot}), are affected in a similar way \citep[see also][]{BildstenChakrabarty2001}. 

\section{Computational model}\label{sec:model}

Having established the important ingredients in spider evolution in Section \ref{sec:general}, we now present our computational model. Specifically, we replace the arbitrary mass-loss rate $\dot{m}$ with a physically motivated model, and we modify the companion star's outer boundary condition to include incident irradiation.

\subsection{Angular momentum and mass loss}\label{sec:mass_loss}

In our model, since evaporation is weak \citepalias{PaperI}, the only way for a companion to lose significant mass is through Roche-lobe overflow. We assume that systems maintain stable Roche-lobe overflow by persistent loss of orbital angular momentum. Under this assumption, the mass-loss rate is governed by the rate at which angular momentum is extracted from the binary system. We consider here two mechanisms to remove angular momentum: gravitational waves and magnetic braking. These mechanisms are important elements in previous calculations of spider evolution \citep{Benvenuto2012, Chen2013}, though the nature of magnetic braking in our scenario is very different.   

\subsubsection{Gravitational waves}\label{sec:GW}

The binary loses orbital angular momentum to gravitational waves on a time-scale \citep{LandauLifshitz1971}
\begin{equation}\label{eq:t_gw}
    t_{\rm GW}=\frac{5}{32}\frac{c^5a^4}{G^3Mm(M+m)},
\end{equation}
where $c$ is the speed of light. We assume a pulsar mass $M=1.4 {\rm M}_{\sun}$, consistent with the ATNF catalogue  \citep{Manchester2005}. In our \textsc{mesa} model, the distance to the pulsar $a$ is calculated from the companion's mass and radius using equation \eqref{eq:rlobe}.
Through stable Roche-lobe overflow, gravitational wave emission leads to mass loss on a similar time-scale $m/\dot{m}=t_{\rm GW}$, up to an uncertain order-unity factor which we omit \citep{Rappaport1982}.

As seen in Fig. \ref{fig:mdot}, gravitational waves do not play a major role in the current evolution of most observed black widows and redbacks (the dashed magenta line marks $t_{\rm GW}=10\textrm{ Gyr}$). Another sink of angular momentum is required to explain how the companions in those systems reached their masses and periods. Nonetheless, as we show in Section \ref{sec:results}, some of our black widow evolutionary tracks pass through a period minimum, where gravitational wave emission accelerates the evolution, before reaching their final configuration. 

\subsubsection{Magnetic braking}\label{sec:braking}

A wind emitted from a spinning magnetized companion is forced to corotate with the companion's magnetic field up to the Alfv\'en radius, which can be much larger than $r$ \citep{WeberDavis67}. Thus, even a weak wind can carry significant angular momentum and spin down the companion. With periods of several hours, the spins of spider companions are tidally locked to their orbits, such that magnetic braking taps into the binary system's orbital angular momentum and pushes it towards Roche-lobe overflow.

Previous studies \citep{Benvenuto2012,Benvenuto2014,Chen2013} employed an empirical magnetic braking prescription that was calibrated to main sequence stars of about a solar mass, with rotation periods of days to weeks \citep{VerbuntZwaan1981,Rappaport1983}. It is not clear how to extrapolate this prescription to spider companions which reach a lower mass, rotate much faster, and are inflated such that they deviate from the main sequence (Fig. \ref{fig:mdot}). Moreover, such classical magnetic braking prescriptions assume that stars power their own winds, irrespective of any binary companion. In our case, however, the host pulsar's incident radiation also drives a wind off the companion star. Although it is too weak to evaporate the companion on a Gyr time-scale \citepalias{PaperI}, this ablated wind is still at least as strong as the measured spontaneous outflows from the sun and other solar-like stars \citep{Wood2002,Wood2005}. 

A key feature in previous black widow and redback calculations is the abrupt shutoff of magnetic braking once the companion's radiative core vanishes and it becomes fully convective, at about $0.3{\rm M}_{\sun}$ on the main sequence \citep{Chen2013,DeVito2020}. This procedure, following \citet{Rappaport1983}, is based on the assumption that the magnetic field is generated at the inner radiative--convective boundary (the `tachocline'). Observations of low-mass stars, however, indicate that these fully convective objects can generate strong kG fields \citep{ReinersBasri2007}. Furthermore, we show in Section \ref{sec:results} that the pulsar's irradiation slows down the companion's cooling, significantly reducing the rate at which its interior releases heat. As a result, strongly irradiated companions retain a radiative core down to much lower masses $\sim 10^{-2}{\rm M}_{\sun}$.

Motivated by the reasons above, we implement here a different magnetic braking prescription, in which the companion's wind, its magnetic field, and the interaction between the two are all evaluated from first principles that are relevant to spider systems. Specifically, in \citetalias{PaperI} we used a hydrodynamical model to calculate the strength of the ablated wind and the rate at which it extracts orbital angular momentum from a Roche-lobe filling companion. The magnetic braking time-scale is given by (re-scaling equation 22 of \citetalias{PaperI}) 
\begin{equation}\label{eq:t_mag}
t_{\rm mag}=2.0\left(\frac{L_{\rm MeV}}{{\rm L}_{\sun}}\right)^{-4/9}\left(\frac{P}{\textrm{h}}\right)^{-34/27}\left(\frac{m}{{\rm M}_{\sun}}\right)^{2/27}\left(\frac{B}{\textrm{kG}}\right)^{-4/3}\textrm{ Gyr},
\end{equation}
where $L_{\rm MeV}$ is the $\sim$MeV $\gamma$-ray luminosity of the pulsar (a portion of its total spin-down power) and $B$ is the companion's magnetic field. As explained in \citetalias{PaperI}, MeV photons deposit their energy high in the companion's atmosphere through Compton scatterings, launching a hydrodynamical wind. Both more energetic and less energetic photons deposit their energy deeper in the atmosphere, where the heated gas can effectively cool and remain below the escape velocity.

We estimate the large-scale magnetic field strength using the \citet{Christensen2009} relation, which assumes an equipartition between the magnetic energy density and the kinetic energy density in the convective eddies $B^2\sim \rho v_{\rm conv}^2$ for rapidly rotating objects (Rossby number ${\rm Ro}\ll1$). 
The convective flow velocity $v_{\rm conv}$ adjusts to transport the companion's internal energy flux $\rho v_{\rm conv}^3=L_{\rm int}/(4\upi r^2)$. The magnetic field is thus given by 
\begin{equation}\label{eq:B}
B=\beta\rho^{1/6}\left(\frac{L_{\rm int}}{4\upi r^2}\right)^{1/3},    
\end{equation}
where $\beta$ is a normalization coefficient; we nominally take $\beta=1$.
This relation fits well both solar system planets and rapidly rotating stars (slowly rotating stars behave differently, and their large-scale magnetic fields may be set by the tachocline). It has also successfully predicted the relatively strong fields of irradiated hot Jupiters \citep{YadavThorngren2017,Cauley2019}, which are in many ways analogous to black widow companions. We note that equation \eqref{eq:B} differs by a factor of 1.25 from the normalization used by \citet{YadavThorngren2017}.
We calculate the internal (i.e. net) cooling luminosity of the companion $L_{\rm int}$ by subtracting the incident power, deposited by the pulsar, from the companion's total luminosity $L$
\begin{equation}\label{eq:L_int}
L_{\rm int}=L-L_{\rm irr}\left(\frac{r}{2a}\right)^2,    
\end{equation}
where $L_{\rm irr}\gtrsim L_{\rm MeV}$ is the pulsar's irradiating luminosity, which sets the companion's outer boundary condition (Section \ref{sec:irradiation}). The evaporative wind is launched primarily by the $\sim$MeV $\gamma$-rays ($L_{\rm MeV}$), as explained in \citetalias{PaperI}, whereas a broader spectrum ($L_{\rm irr}$) is deposited deeper in the atmosphere. We note that $L_{\rm irr}$ itself is likely smaller than the total spin-down power, as some of that power is carried away by a low frequency Poynting flux which might not directly interact with the companion's atmosphere. When the companion star is massive, it is luminous enough such that $L_{\rm int}\approx L$. As the companion loses mass and its own internal luminosity decreases, only a fraction of the total luminosity $L_{\rm int}\ll L$ is delivered by convection from the interior (setting the magnetic field), with the rest being re-radiated by the atmosphere.

We sum the effects of gravitational waves and magnetic braking by removing mass at a rate $\dot{m}$, given by
\begin{equation}\label{eq:mdot_sum}
\frac{\dot{m}}{m}=\frac{1}{t_{\rm GW}}+\frac{1}{t_{\rm mag}}.
\end{equation}
We emphasize that unlike previous studies, we assume in our nominal runs that magnetic braking persists even after the companion's radiative core vanishes. This is motivated by both theory \citep[][and references therein]{Christensen2009} and observations of low-mass stars and planets \citep{ReinersBasri2007,Cauley2019}, which suggest that strong magnetic fields can be generated in the absence of a tachocline. For completeness, in Section \ref{sec:results} we also test a scenario in which the magnetic field is reduced once the companion's interior becomes fully convective.

\subsection{Irradiation}\label{sec:irradiation}

We demonstrated in Section \ref{sec:general} that Roche-lobe filling (including almost filling) black widow and redback companions can survive for longer than $10^7-10^8$ yr with large radii (above the main sequence or degenerate value for their mass) only if their cooling is slowed down by the pulsar's irradiation. Specifically, the boundary conditions on the night side of the tidally locked companion, which cools faster than the irradiated day side, are the ones that determine the cooling rate. If the night-side boundary conditions are unchanged, the cooling time $t_{\rm KH}$ increases by merely a factor of 2 as heat leaks unimpeded through one hemisphere. To increase $t_{\rm KH}$ further, the absorbed pulsar radiation must be redistributed over both hemispheres by atmospheric winds, similarly to hot Jupiters \citep{GuillotShowman2002,ShowmanGuillot2002}. Such heat redistribution also seems to be favoured by fitting to optical light curves of spider companions \citep{KandelRomani2020,Voisin2020}.

We implement irradiation in \textsc{mesa} using the $F_\star$--$\Sigma_\star$ method, where $F_\star$ is the deposited flux and $\Sigma_\star$ is the deposition mass column density \citep{Paxton2013}. We parametrize the flux as
\begin{equation}\label{eq:F_star}
F_\star=\frac{L_{\rm irr}}{4\upi a^2},    
\end{equation}
where $L_{\rm irr}$ represents the pulsar's irradiating luminosity. As discussed above, $L_{\rm irr}$ is plausibly somewhat larger than the $\gamma$-ray luminosity $L_{\rm MeV}$ (which drives the evaporative wind) as it covers a broader part of the pulsar's spectrum. We treat $L_{\rm MeV}$ and $L_{\rm irr}$ as separate free parameters --- in fact our only free parameters --- to allow for different pulsar spectra (i.e. different $L_{\rm MeV}/L_{\rm irr}$), or different efficiencies of winds at redistributing the energy. As we demonstrate in Section \ref{sec:optical}, $L_{\rm irr}$ is directly linked to the companion's night-side temperature, which can be inferred from optical observations.

The effective night-side deposition depth is determined by comparing the advection and diffusion time-scales, similarly to \citet{GinzburgSari2016}. We omit order-unity coefficients due to the complex multi-dimensional nature of the atmospheric flows and since our results are highly insensitive to $\Sigma_\star$. The day--night temperature difference powers atmospheric winds, which advect the deposited heat from the day side to the cooler night side. Deep enough in the atmosphere, the photon diffusion time exceeds the advection time $t_{\rm adv}=r/v_{\rm adv}$, allowing the two hemispheres to reach thermal equilibrium with each other before the excess heat escapes outwards. The companion's atmosphere is in the low Rossby number regime ($v_{\rm adv}\ll\Omega r$, as we show below), implying that the advection velocity $v_{\rm adv}$ is set by balancing the thermal forcing with the Coriolis force: $c_{\rm s}^2/r\sim \Omega v_{\rm adv}$, where $\Omega\equiv 2\upi/P$ (at $P$ of several hours, spider companions are tidally locked such that both the orbital and rotational periods are equal to $P$) and $c_{\rm s}$ is the speed of sound \citep{ShowmanGuillot2002}; note that at the deposition depth, the day--night temperature ratio is by definition of order unity (the temperature difference and the winds die out with depth). The advection velocity at the deposition depth is therefore 
\begin{equation}
    v_{\rm adv}=\frac{c_{\rm s}^2}{\Omega r}=\frac{c_{\rm s}^2}{v_{\rm esc}},
\end{equation}
where $\Omega r$ is similar to the escape velocity $v_{\rm esc}\equiv(Gm/r)^{1/2}$ for tidally locked Roche-lobe filling companions. As we show below, the night-side temperature is $\sim 3\times 10^3\textrm{ K}$, such that $c_{\rm s}\ll v_{\rm esc}$ and consequently $v_{\rm adv}\ll c_{\rm s}\ll v_{\rm esc}\sim\Omega r$, justifying the low Rossby number.

\begin{figure}
\includegraphics[width=\columnwidth]{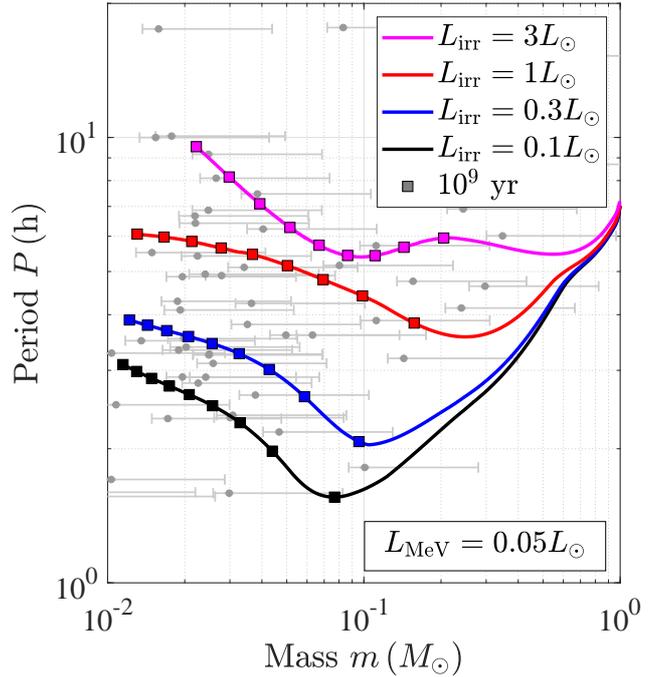}
\caption{Black widow evolutionary tracks that follow the mass and orbital period of an initially 1 Gyr old, $1 {\rm M}_{\sun}$ main sequence companion up to an age of 10 Gyr (square markers are separated by 1 Gyr). Roche-lobe overflow is assumed throughout the evolution to connect the companion's radius to the orbital period. Mass is removed at the same rate as orbital angular momentum, which is lost to both gravitational waves and magnetic braking. The latter is given by the interaction of the companion's magnetic field, powered by convection \citep[][our equation \eqref{eq:B} with the nominal $\beta=1$]{Christensen2009}, with the ablated wind launched by the pulsar's MeV $\gamma$-ray luminosity $L_{\rm MeV}$ ($=0.05 {\rm L}_{\sun}$ for all tracks). A broader spectrum of the pulsar's luminosity $L_{\rm irr}\gtrsim L_{\rm MeV}$ deposits heat deep in the companion's atmosphere and thereby slows its cooling. Higher values of $L_{\rm irr}$ lead to longer-period black widows (the legend and the figure have the same order). The observations (grey dots) are taken from Fig. \ref{fig:mdot}.}
\label{fig:tracks}
\end{figure}

\begin{figure}
\includegraphics[width=\columnwidth]{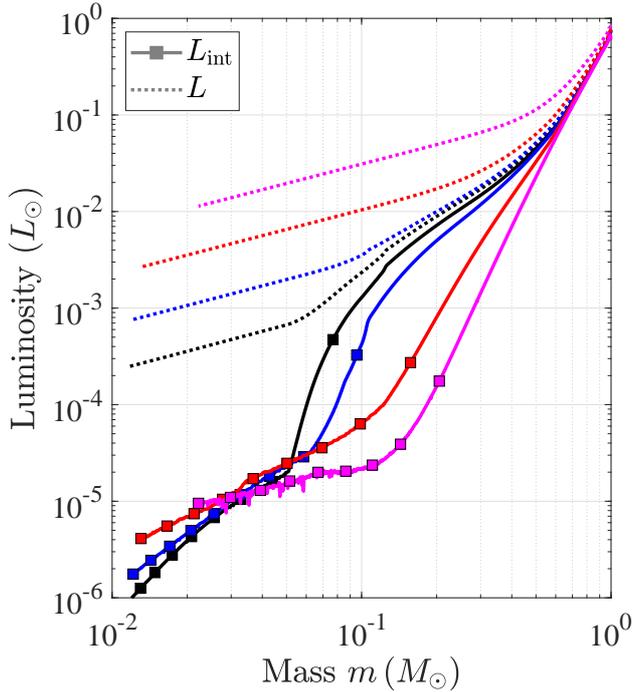}
\caption{The companion's total luminosity $L$, and its internal luminosity $L_{\rm int}=L-L_{\rm irr}[r/(2a)]^2$ for our nominal tracks (Fig. \ref{fig:tracks}, same line colours). At high masses, the companion star is self-luminous and $L_{\rm int}\sim L$. At lower masses, most of the companion's luminosity is re-radiated pulsar energy $L\approx L_{\rm irr}[r/(2a)]^2\propto L_{\rm irr}m^{2/3}$, and only a small fraction $L_{\rm int}\ll L$ is delivered from the convective interior. A stronger pulsar irradiation $L_{\rm irr}$ reduces $L_{\rm int}/L$ (see the analytical analysis in Section \ref{sec:analytical}).}
\label{fig:Lint}
\end{figure}

\begin{figure}
\includegraphics[width=\columnwidth]{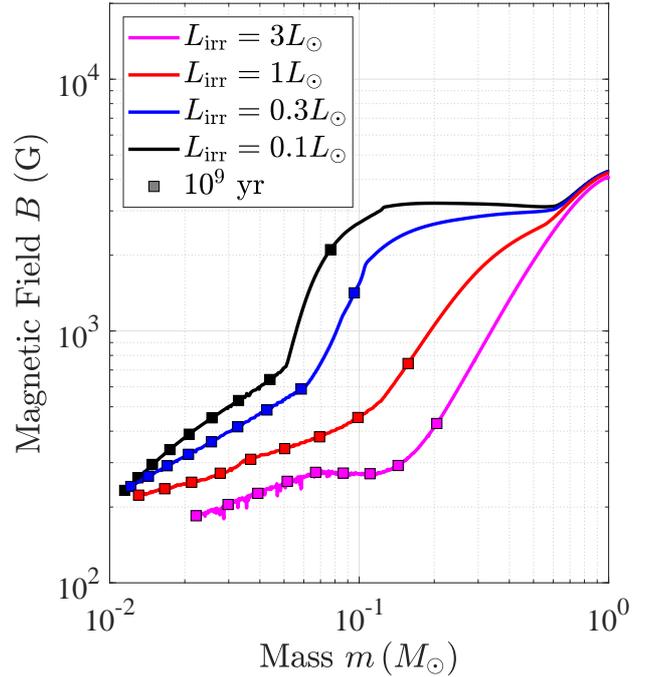}
\caption{The companion's magnetic field $B$ for our nominal tracks (Fig. \ref{fig:tracks}), computed using the \citet{Christensen2009} relation, given by our equation \eqref{eq:B}. This relation postulates an equipartition between the magnetic energy and the kinetic energy in convective eddies, and thus relates $B$ to the internal convective luminosity $L_{\rm int}$ (Fig. \ref{fig:Lint}).}
\label{fig:B}
\end{figure}

The time it takes a layer of mass $\Delta m=4\upi r^2\Sigma_\star$ and optical depth $\tau$ to radiate away its thermal energy $\sim\Delta m c_{\rm s}^2$ by diffusion is
\begin{equation}\label{eq:t_rad}
    t_{\rm rad}=\frac{\Delta m c_{\rm s}^2\tau}{4\upi r^2 \sigma T^4}=\frac{\Sigma_\star\tau c_{\rm s}^2}{F_\star},
\end{equation}
which reduces to equation (29) of \citet{GinzburgSari2016} for a constant opacity. The second equality in equation \eqref{eq:t_rad} holds for the deposition level, which is heated to a temperature $T\sim(F_\star/\sigma)^{1/4}$ whenever irradiation is important (see Section \ref{sec:optical}; $\sigma$ is the Stefan--Boltzmann constant).
We find the deposition depth using the condition $t_{\rm rad}=t_{\rm adv}=r/v_{\rm adv}$:
\begin{equation}\label{eq:depth}
\Sigma_\star\tau=\frac{rF_\star}{c_{\rm s}^2v_{\rm adv}}=\frac{rv_{\rm esc}F_\star}{c_{\rm s}^4}.    
\end{equation}
We change the deposition depth $\Sigma_\star$ in \textsc{mesa} every time-step such that $\Sigma_\star \tau(\Sigma_\star)$ satisfies equation \eqref{eq:depth}, with $c_{\rm s}$ estimated at the surface \citep[the radiative layer induced by irradiation is isothermal up to an order unity factor, such that $c_{\rm s}$ is roughly uniform up to the deposition level; e.g.][]{ArrasBildsten2006}. Our results depend only weakly on this choice of $\Sigma_\star$: we find that an order of magnitude difference in $\Sigma_\star$ changes black widow periods and masses by less than 50 per cent, and the qualitative behaviour remains the same.

$F_\star$ is redistributed effectively from the irradiated day side to the night side only if the day-side deposition is deeper than $\Sigma_\star$ that is calculated by equation \eqref{eq:depth}; otherwise most of the heat diffuses out before being advected between the hemispheres. For typical black widow companions, equation \eqref{eq:depth} gives $\Sigma_\star\sim 10^2\textrm{ g cm}^{-2}$. The day-side energy deposition depth is given by the cross section to pair production, which is the dominant interaction for the energetic GeV photons observed by {\it Fermi} for black widow host pulsars \citep{Abdo2013}. Using the \citet{BetheHeitler34} formula, we find that the day-side $\Sigma_\star$ for pair production is larger by a factor of a few compared to equation \eqref{eq:depth} --- justifying our treatment of the irradiation.

\section{Results}\label{sec:results}

Our model, described in Section \ref{sec:model}, has two free parameters: $L_{\rm MeV}$ and $L_{\rm irr}\gtrsim L_{\rm MeV}$, which represent the pulsar's $\gamma$-ray (specifically, $\sim$MeV) and broader spectrum luminosities, respectively. In principle, the two parameters entangle together non-trivially to determine the pulsar companion's evolution. The companion's mass-loss rate $\dot{m}$ (dictated by equation \ref{eq:mdot_sum}) is a function of $L_{\rm MeV}$ (which drives the evaporative wind), but also of the magnetic field $B$, which is powered by the companion's internal luminosity $L_{\rm int}$ --- itself a function of the irradiation $L_{\rm irr}$. For a given $\dot{m}$, the companion's period evolves to satisfy $t_{\rm KH}(L_{\rm irr},P)\sim m/\dot{m}(L_{\rm MeV}, L_{\rm irr}, P)$, as demonstrated in Section \ref{sec:general}. We emphasize that the companion's cooling time $t_{\rm KH}$ depends critically on the irradiation of its surface by $L_{\rm irr}$, since the latter significantly reduces $L_{\rm int}$.
In practice, we find that $L_{\rm MeV}$ and $L_{\rm irr}$ conveniently map to the black widow observables: $L_{\rm irr}$ approximately determines the orbital period reached by an evolutionary track, whereas $L_{\rm MeV}$ sets the pace at which the companion loses its mass along the track. We explain this analytically in Section \ref{sec:analytical}. 

\begin{figure}
\includegraphics[width=\columnwidth]{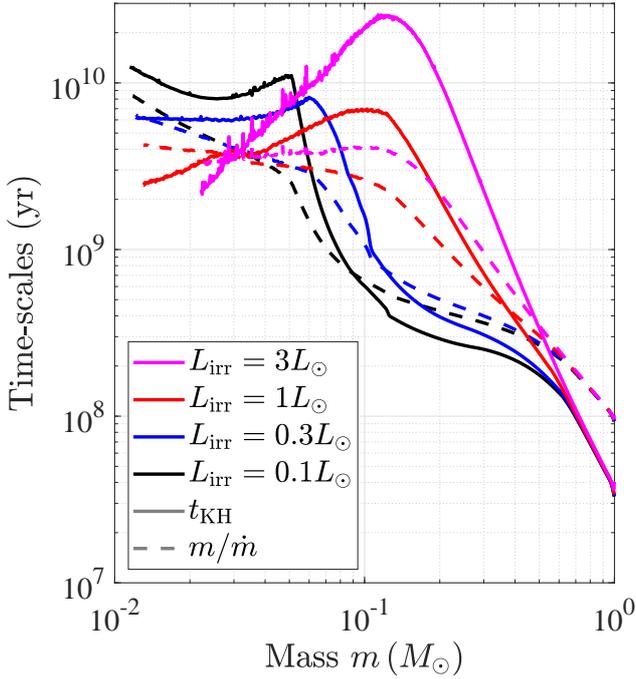}
\caption{The companion's Kelvin--Helmholtz cooling time $t_{\rm KH}$ (solid lines) and its mass-loss time $m/\dot{m}$ (dashed lines) for our nominal tracks (Fig. \ref{fig:tracks}, same line colours). $t_{\rm KH}$ is given by the ratio of the companion's thermal energy to its internal luminosity $L_{\rm int}$. $\dot{m}$ is given by magnetic braking and gravitational waves, which enforce Roche-lobe overflow; equation \eqref{eq:mdot_sum}. The tracks evolve such that $t_{\rm KH}\sim m/{\dot m}$. When the companions are reduced to black widow masses, their evolution times are several Gyr (when pulsar irradiation is ignored, black widow life-times are orders of magnitude shorter; see Fig. \ref{fig:tcool}).}
\label{fig:timescales}
\end{figure}

\begin{figure}
\includegraphics[width=\columnwidth]{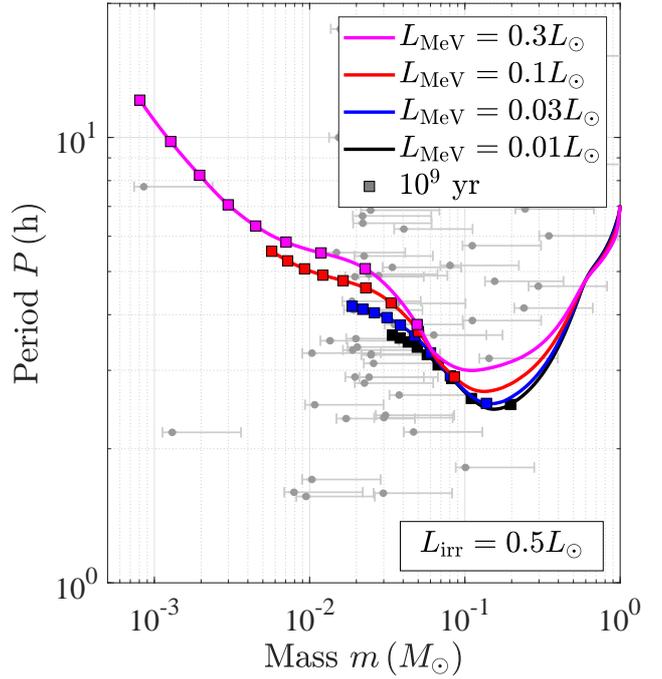}
\caption{Similar to Fig. \ref{fig:tracks} (with the same nominal $\beta=1$), but showing the dependence on the pulsar's MeV $\gamma$-ray luminosity $L_{\rm MeV}$ for a constant broader spectrum irradiation $L_{\rm irr}=0.5{\rm L}_{\sun}\gtrsim L_{\rm MeV}$. This and Fig. \ref{fig:tracks} demonstrate that $L_{\rm MeV}$ controls the speed at which companions lose mass, whereas $L_{\rm irr}$ controls their orbital period.}
\label{fig:Lgamma}
\end{figure}

\begin{figure}
\includegraphics[width=\columnwidth]{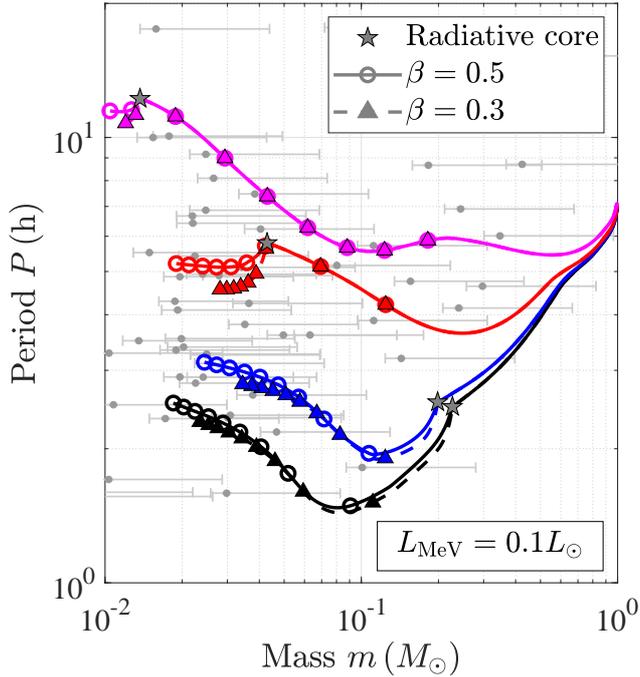}
\caption{Same as Fig. \ref{fig:tracks}, but for $L_{\rm MeV}=0.1{\rm L}_{\sun}$ (for all tracks and at all times), and with the companion's magnetic field $B$ reduced by an arbitrary factor ($\beta=0.5$ or 0.3), with respect to equation \eqref{eq:B}, once the radiative core vanishes (marked with grey stars). The other markers in each track are separated by 1 Gyr. We consider the radiative core vanished once its mass fraction drops below $10^{-4}$ of the total mass $m$. This mass fraction drops abruptly, such that the tracks are not sensitive to the threshold value. Companions subject to strong irradiation $L_{\rm irr}$ emit less internal flux, allowing their cores to remain radiative down to a lower $m$. This effect, combined with gravitational waves at short periods, enables the tracks to reach black widow masses in 10 Gyr, even with weaker magnetic braking. Having said that, we choose a higher $L_{\rm MeV}=0.1 {\rm L}_{\sun}$ to slightly expedite the evolution while keeping $L_{\rm MeV}\leq L_{\rm irr}$.}
\label{fig:convective}
\end{figure}

\begin{figure}
\includegraphics[width=\columnwidth]{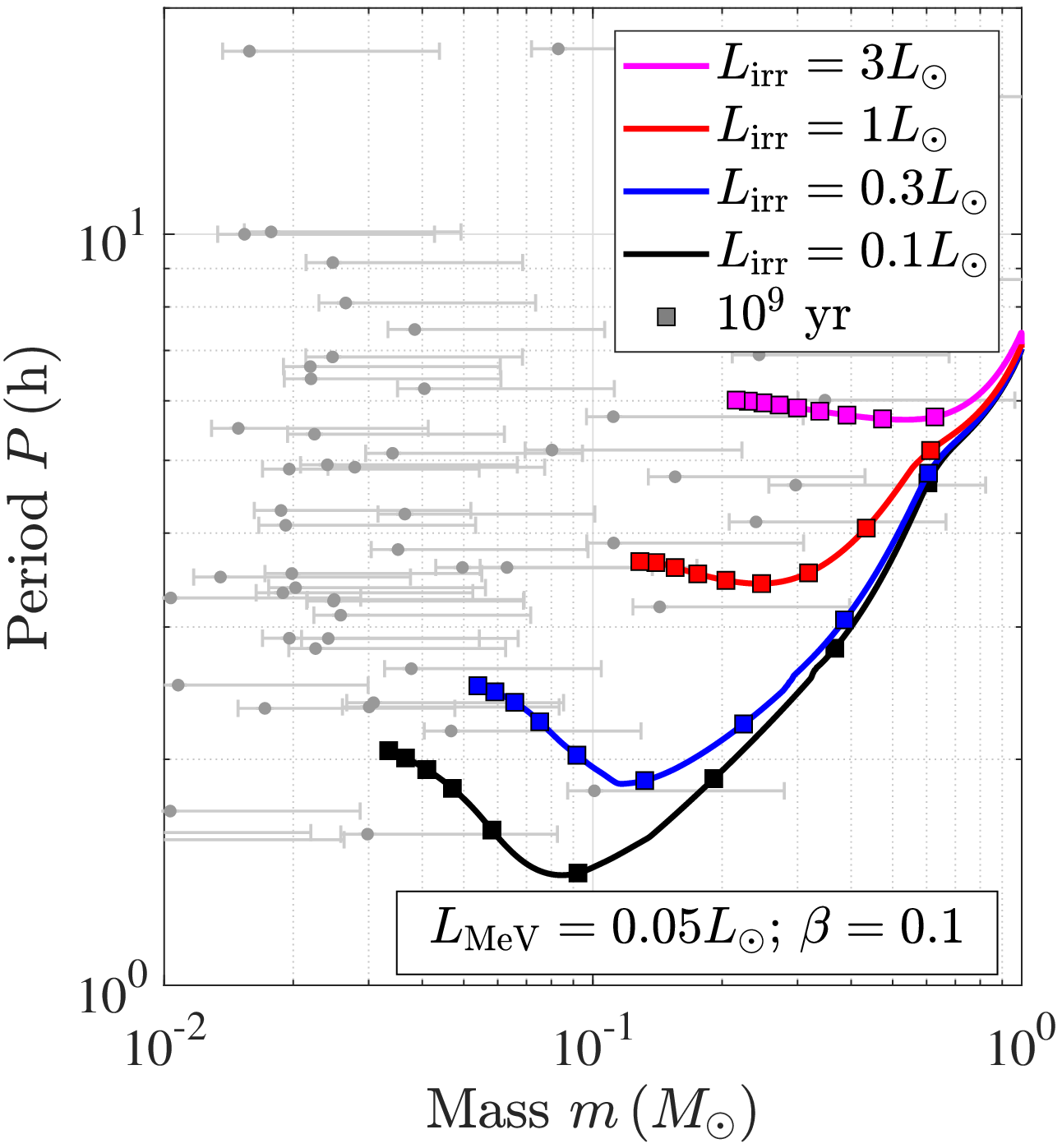}
\caption{Same as Fig. \ref{fig:tracks}, but the companion's magnetic field $B$ is smaller ($\beta=0.1$) than our nominal equation \eqref{eq:B} at all times (i.e. regardless of a radiative core). Companions with weak fields evolve more slowly, due to the weaker magnetic braking, and remain redbacks if subject to strong irradiation $L_{\rm irr}$. If $L_{\rm irr}$ is low, companions evolve to shorter periods, where gravitational waves may transform them into black widows. A bimodal population of companion $B$ fields --- one follows the evolution in Fig. \ref{fig:tracks} and one follows that shown here --- could explain the tentative evidence for a bimodal distribution of companion masses.}
\label{fig:weak}
\end{figure}

With this intuition, we compare our model to the observed black widow population in Fig. \ref{fig:tracks}. There are almost no black widow companions below $10^{-2} {\rm M}_{\sun}$, although these could have been detected \citep{WolszczanFrail92, Bailes2011,Spiewak2018}. We interpret this sharp edge in the black widow mass distribution (Fig. \ref{fig:mdot}) as a maximum system age. Assuming that no black widow is older than about 10 Gyr sets $L_{\rm MeV}\sim 0.05 {\rm L}_{\sun}$, though this value is degenerate with the magnetic field normalization $\beta$; see Section \ref{sec:analytical}. In order to reproduce the observed black widow periods, which span about 2--10 h, we vary $L_{\rm irr}$ between 0.1 and 3 ${\rm L}_{\sun}$. It is reassuring to find that the $L_{\rm irr}$ and $L_{\rm MeV}$ values that reproduce the observed black widow population in the $m$--$P$ plane satisfy $L_{\rm irr}\gtrsim L_{\rm MeV}$, since the two parameters are determined from the mass and period distributions independently of each other. We are also encouraged that measured high-energy luminosities of black widow pulsars span the same $0.1-3\,{\rm L}_{\sun}$ range \citep{Abdo2013}, although {\it Fermi} is sensitive only to 0.1--100 GeV photons.\footnote{In \citet{Abdo2013} $L_\gamma$ refers to 0.1--100 GeV photons, whereas in \citetalias{PaperI} $L_\gamma$ refers to $\sim$MeV photons. Here, we use $L_{\rm MeV}$ to avoid confusion.} 

The companion's $L_{\rm int}$ and $B$ along the tracks in Fig. \ref{fig:tracks} are plotted in Figs \ref{fig:Lint} and \ref{fig:B}.
As Fig. \ref{fig:Lint} shows, the pulsar's irradiation inhibits the companion's internal luminosity $L_{\rm int}$, thereby lengthening its Kelvin--Helmholtz cooling time $t_{\rm KH}$. The tracks evolve such that $t_{\rm KH}\sim m/\dot{m}$, with both time-scales equal to several Gyr for black widows (Fig. \ref{fig:timescales}; compare with Fig. \ref{fig:tcool}, which disregards the pulsar's irradiation). We emphasize that this condition is not imposed in the simulation; instead, the mass-loss rate $\dot{m}$ is dictated by magnetic braking and gravitational waves according to equation \eqref{eq:mdot_sum}.

Fig. \ref{fig:Lgamma} complements Fig. \ref{fig:tracks} by showing the dependence on our second free parameter $L_{\rm MeV}$. As also found analytically in Section \ref{sec:analytical}, the orbital period $P(m)$ is insensitive to $L_{\rm MeV}$, whose main effect is setting the pace at which companions lose mass. A very high $L_{\rm MeV}$ (but still $\lesssim L_{\rm irr}$), or more likely (because the pulsar's spin-down power in that system is only $\approx 0.1{\rm L}_{\sun}$), a strong magnetic field $\beta>1$, might explain the extremely low mass companion to PSR J2322$-$2650 \citep[][]{Spiewak2018}. PSR J1719$-$1438 \citep{Bailes2011} is more difficult to explain due to the combination of a low mass and a short orbital period.

The simulated systems in Fig. \ref{fig:tracks} evolve quickly (in about a Gyr) into redbacks, and after that into black widows, reasonably reproducing the bulk of the black widow population after several Gyr. While we also reproduce the short-period redback population, our nominal tracks do not reach the longer period redbacks (Fig. \ref{fig:mdot}); these require $L_{\rm irr}>10 {\rm L}_{\sun}$ in the framework of our model, which would imply a roughly 100 per cent efficiency of converting the pulsar's spin-down power into $L_{\rm irr}$. Another mechanism to power such high $L_{\rm irr}$ is by accretion luminosity onto the pulsar, which may be strong enough for companions with redback masses \citep{Ruderman89E,Benvenuto2012}.

It is instructive to compare our evolutionary tracks to previous studies that computed similar tracks \citep{Chen2013,Benvenuto2015MNRAS, DeVito2020}. In those studies, companions are also initially kept at stable Roche-lobe overflow, but the angular momentum loss rate is calculated using an extrapolation of the \citet{Rappaport1983} formula, whereas we derive an appropriate rate for evaporating spiders (Section \ref{sec:braking}). Following the \citet{Rappaport1983} prescription, previous studies shut off magnetic braking when the companion becomes fully convective ($m\simeq 0.3 {\rm M}_{\sun}$) and at the same time turn on evaporation by the pulsar with an arbitrary efficiency. The companion detaches from its Roche lobe and the orbit expands on a track that depends on the evaporation efficiency \citep[fig. 4 in][]{Chen2013}: high efficiencies lead to redbacks whereas low efficiencies, combined with angular momentum loss to gravitational waves, form black widows. We find, on the other hand, that evaporation on its own is too weak to evolve the system on a Gyr time-scale \citepalias{PaperI}. Instead, the evaporative wind couples to the companion's magnetic field and magnetic braking maintains stable Roche-lobe overflow even at late times. One notable difference in the resulting tracks is that many of our simulated black widows are descendent from more massive redbacks. In previous studies, the two spider groups share a common ancestor, but evolve on separate branches. Another difference is the Roche-lobe filling factor: In our scenario, sustained magnetic braking keeps the companion at or near Roche-lobe overflow. In previous studies, on the other hand, strong evaporation and the lack of sufficient angular momentum sinks allow the companion to detach and significantly under-fill its Roche lobe. Analysis of optical black widow light curves suggests that the companions generally fill or almost fill their Roche lobes, with filling factors (ratio of the companion's radius to that of the Roche lobe) $\gtrsim 0.7$ for 8 out of 9 systems \citep{Draghis2019}. All the redback systems considered by \citet{DeVito2020} also have filling factors of 0.7 or above. 
The generally high observed filling factors seem consistent with our sustained magnetic braking scenario.
We note that according to \citetalias{PaperI} (specifically table 2), a small minority of black widows may evaporate their companions directly faster than magnetic braking can push them towards Roche-lobe overflow (J0024$-$7204P, J1701$-$3006E/F). Such systems could have lower filling factors.    

For further comparison with previous studies, we follow the radiative core of the companion in Fig. \ref{fig:convective}. Companions subject to weak irradiation remain close to the main sequence and lose their radiative cores at a mass $m\simeq 0.2 {\rm M}_{\sun}$. Stronger irradiation $L_{\rm irr}$ significantly reduces the companion's internal luminosity $L_{\rm int}$. With less heat leaving the companion's interior, its core remains radiative down to much lower companion masses $m\sim 10^{-2} {\rm M}_{\sun}$. In Fig. \ref{fig:convective} we also mimic the \citet{Rappaport1983} prescription, which assumes that magnetic fields are generated at the tachocline, and weaken the companion's magnetic field $B$ once the radiative core vanishes. Specifically, we reduce $B$ by an order-unity factor that is large enough to explain the period gap in cataclysmic variables \citep[][this is one of the original motivations for such a prescription]{SpruitRitter1983}.
Our results are largely unchanged: At low $L_{\rm irr}$, companions evolve to short periods, where gravitational waves extract sufficient angular momentum. At higher $L_{\rm irr}$, radiative cores, and therefore strong magnetic braking, persist down to much smaller $m$. In both cases, our evolutionary tracks produce black widows after several Gyr. This variation of our model might fit the observed spider population slightly better than our nominal tracks (Fig. \ref{fig:tracks}). The prolonged evolution time-scales at low $m$, when compared to higher masses, potentially explain why black widows are somewhat more common than redbacks --- spiders spend more time at lower masses. 
We note that shutting off $B$ entirely (i.e. $\beta=0$) when the radiative core is gone would not be able to explain low-mass companions ($< 4\times 10^{-2}{\rm M}_{\sun}$) at most orbital periods. For example, the $L_{\rm irr}=1{\rm L}_{\sun}$ track (red line) in Fig. \ref{fig:convective} would not evolve much beyond the grey star marker in that case.

\begin{figure}
\includegraphics[width=\columnwidth]{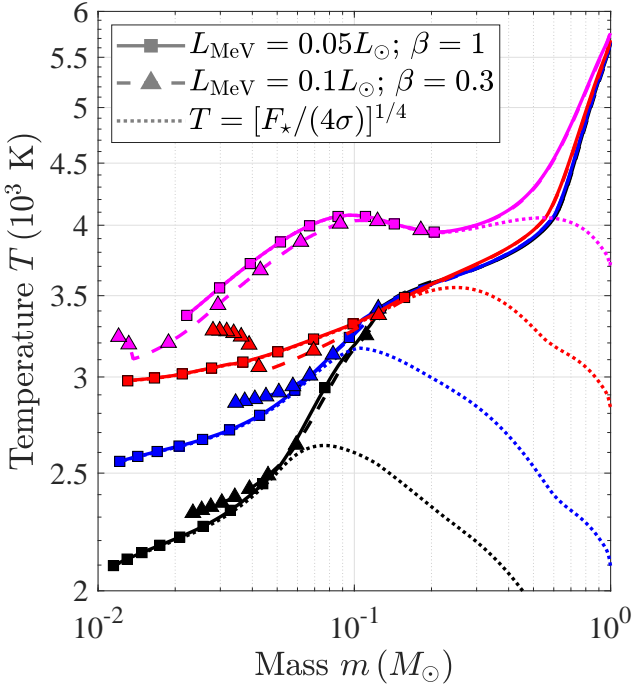}
\caption{Effective temperature of the simulated companions presented in Figs \ref{fig:tracks} and \ref{fig:convective} (solid and dashed lines; same colours, markers, and order). At high masses, the temperature follows that of a main sequence isolated star. As the companion loses mass, the temperature decreases until it approaches the irradiation temperature $T\propto L_{\rm irr}^{1/4}P^{-1/3}$ (dotted lines, plotted only for the $L_{\rm MeV}=0.05 {\rm L}_{\sun}$ case), which it then follows at low masses.}
\label{fig:Tnight}
\end{figure}

\begin{figure}
\includegraphics[width=\columnwidth]{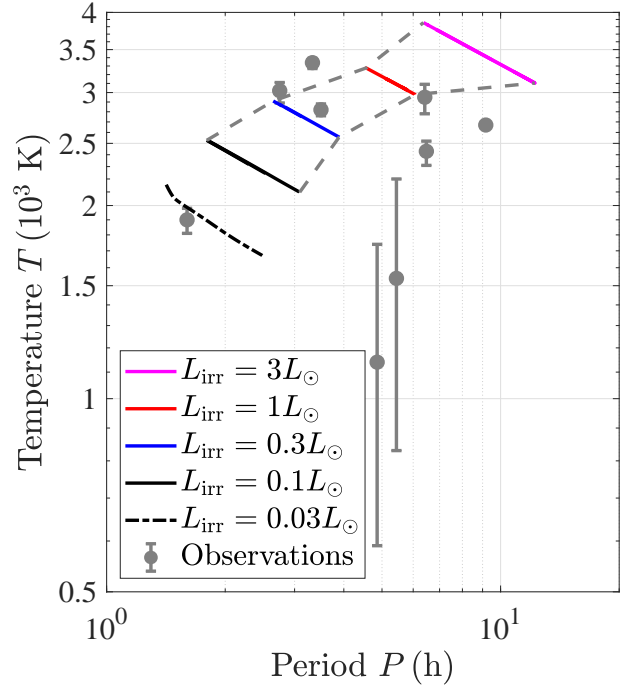}
\caption{Observed night-side temperatures of black widow companions, inferred from their optical light curves by \citet{Draghis2019}. The solid lines indicate the periods and temperatures reached by our models in Fig. \ref{fig:Tnight} for masses $m<0.05 {\rm M}_{\sun}$ (i.e. the black widow region). The colours and order of these lines are the same as in Figs \ref{fig:tracks} and \ref{fig:Tnight}. The solid lines are connected by dashed grey lines, which approximately bound the region in the period--temperature diagram that our simulated black widows inhabit. We also add a model with $L_{\rm MeV}=L_{\rm irr}=0.03{\rm L}_{\sun}$ (dot--dashed black line).}
\label{fig:Draghis}
\end{figure}

Previous studies identified a paucity of companions with masses in the range 0.05--0.1 ${\rm M}_{\sun}$, separating the lighter black widows from the more massive redbacks \citep{Roberts2013,Chen2013}. As seen in Fig. \ref{fig:mdot}, our updated sample does not feature such a prominent valley in the mass distribution, given the large mass uncertainty. Our nominal model does not reproduce this bimodality either --- redbacks evolve into black widows at a steady pace (Fig. \ref{fig:tracks}). While our model with suppressed magnetic braking after the radiative core disappears predicts a concentration of black widow systems (Fig. \ref{fig:convective}), it does not produce a similar concentration of redbacks, so this model too lacks a mass gap.
Notwithstanding the uncertain statistical significance of this gap, we note that it can be reproduced with a bimodal distribution of magnetic fields.
In Fig. \ref{fig:weak} we present evolutionary tracks of weakly magnetized companions, with magnetic fields smaller by an order of magnitude than our nominal equation \eqref{eq:B}. These tracks lead to the formation of redbacks, and do not evolve into black widows, except for low values of $L_{\rm irr}$. A population of such weakly magnetized companions can therefore cluster at redback masses, reproducing the valley between redbacks and strongly magnetized black widows. The existence of two populations with different magnetic fields is motivated by measurements of stellar rotation periods for FGK stars, which show a bimodality in young clusters \citep{Meibom2011}: Some stars maintain fast rotations of less than a day for several $10^8$ yr, while others spin down to $\sim$10 day periods at the same age, presumably due to magnetic braking; see also \citet{Newton2016} for fully convective M dwarfs, where the ages are less constrained and estimated from Galactic kinematics. A possible interpretation of these rotation observations is a sharp and perhaps stochastic change in the operation of the magnetic dynamo in these stars \citep{Brown2014,Garraffo2015}. The tentative spider mass gap and the stellar rotation period gap thus might be related.

\subsection{Optical light curves}\label{sec:optical}

Optical observations of black widow companions can prove valuable in distinguishing between different formation scenarios. We have already discussed above how the Roche-lobe filling factor, which is constrained by such observations, indicates the relative roles of evaporation and Roche-lobe overflow. Another useful quantity that can be extracted from the optical light curve is the companion's night-side temperature. Since our one-dimensional \textsc{mesa} calculation essentially models the night side (Section \ref{sec:irradiation}), the effective temperature of the simulated companions can be directly compared with this observable. By examining the night-side temperature, we bypass unknown efficiencies in converting the pulsar's spin-down power to $L_{\rm irr}$ and in the transport of energy from the companion's day side to its night side. In our models, the night-side temperature directly sets the companion's cooling rate, and therefore its period evolution along the computed tracks.  

We present the effective temperature along our evolutionary tracks in Fig. \ref{fig:Tnight}. When the companion is massive and luminous, its thermal structure is unaffected by the pulsar's irradiation. As the companion loses mass, its luminosity drops, and the surface temperature becomes dominated by the external irradiation rather than internal cooling. At low masses, the temperature is given by $T\simeq [F_\star/(4\sigma)]^{1/4}\propto L_{\rm irr}^{1/4}P^{-1/3}$, where $\sigma$ is the Stefan--Boltzmann constant and $F_\star$ given by equation \eqref{eq:F_star}. Strongly irradiated companions exhibit higher temperatures, but the higher $L_{\rm irr}$ is partially compensated by the longer periods to which such companions evolve (Fig. \ref{fig:tracks}). As a result, Fig. \ref{fig:Tnight} shows that our models predict that black widow companions have similar night-side temperatures $T\sim 3\times 10^3\textrm{ K}$. 

In Fig. \ref{fig:Draghis} we compare our calculated temperatures with the observed night-side temperatures of black widow companions \citep{Draghis2019}. Our model predicts a correlation between the period and the temperature --- both increase with $L_{\rm irr}$. However, as explained above, the correlation is weak and there is significant overlap in our predicted temperatures for black widow companions at different periods --- the temperature is around $3\times 10^3\textrm{ K}$ for a wide range of periods. Most of the observed black widow companions with well-constrained temperatures are close to our predicted values. We find this encouraging, since our range of $L_{\rm irr}$ was not chosen to fit the optical observations, but rather the periods reached by evolving companions (Fig. \ref{fig:tracks}). The fact that the observed temperatures are similar to those that satisfy $t_{\rm KH}(P,T)\sim m/\dot{m}(P,T)$ provides an independent test of, and some additional confidence in, our mass-loss scheme.

\citet{RomaniSanchez2016} modelled optical light curves with an `intra-binary shock' --- the result of a collision between the pulsar and companion winds --- which reprocesses the pulsar's illumination and heats the companion non-uniformly. In principle, the geometry of the shock and hence the companion's ablated wind strength can thus be directly constrained from the shape of the light curve.
While the mass-loss rates estimated this way are degenerate with the other parameters of the fit, they nonetheless indicate that at least some pulsars may directly (i.e. even without magnetic braking) evaporate their companions in less than a Gyr. 

\subsection{Analytical analysis}\label{sec:analytical}

It is useful to reproduce the evolutionary tracks in Fig. \ref{fig:tracks} with analytical scaling relations. As explained in Section \ref{sec:general}, these tracks are given by $t_{\rm KH} \sim t_{\rm mag}$, where gravitational waves can be neglected, for the most part, in equation \eqref{eq:mdot_sum}. We focus here on the evolution at low masses $m\lesssim 10^{-1}{\rm M}_{\sun}$, once the companion's effective temperature becomes dominated by the pulsar's irradiation (Fig. \ref{fig:Tnight}). Despite the low-mass objects we are considering, gas pressure dominates over degeneracy pressure. This is because irradiation keeps the radius inflated --- larger than that of a zero-temperature object (see Fig. \ref{fig:mdot}). 

The Kelvin--Helmholtz cooling time $t_{\rm KH}$ is given by an irradiated version of the Hayashi-line calculation \citep[e.g.][]{Hansen2004,Kippenhahn2012}, similarly to \citet{ArrasBildsten2006}. We assume an adiabatic index $\gamma=5/3$ and an ${\rm H}^-$ opacity $\kappa\propto p^{1/2}T^{17/2}$, which is relevant for $2000\textrm{ K}\lesssim T\lesssim 10^4\textrm{ K}$. The pulsar's irradiation induces an outer radiative layer at the irradiation temperature $T\propto L_{\rm irr}^{1/4}P^{-1/3}$; the layer is isothermal to within an order unity factor. The radiative layer connects to the companion's (mostly or fully) convective interior at the outer radiative--convective boundary, located at a pressure $p_{\rm rcb}$, which is given by
\begin{equation}
\frac{p_{\rm rcb}}{p_0}=\left(\frac{T}{T_0}\right)^{\gamma/(\gamma-1)}=\left(\frac{T}{T_0}\right)^{5/2},    
\end{equation}
where $p_0\propto m^2r^{-4}$ and $T_0\propto mr^{-1}$ are the central pressure and temperature, respectively; therefore, $p_{\rm rcb}\propto T^{5/2}m^{-1/2}r^{-3/2}$. The optical depth of the radiative layer is given by $\tau_{\rm rcb}\sim(\kappa/g)p_{\rm rcb}$, where $g\equiv Gmr^{-2}$. The companion's internal luminosity is given by diffusion through the radiative layer
\begin{equation}\label{eq:L_int_anal}
L_{\rm int}\sim\frac{4\upi r^2\sigma T^4}{\tau_{\rm rcb}}\propto m^{7/4}r^{9/4}T^{-33/4}\propto m^{5/2}P^{17/4}L_{\rm irr}^{-33/16},   
\end{equation}
where the last proportionality is for Roche-lobe filling companions, which satisfy $r\propto m^{1/3}P^{2/3}$ according to equation \eqref{eq:period_r}. Equation \eqref{eq:L_int_anal} indicates that strong irradiation suppresses the companion's cooling luminosity. The cooling time for non-degenerate companions is
\begin{equation}\label{eq:t_cool_anal}
t_{\rm KH}\sim \frac{Gm^2}{rL_{\rm int}}\propto m^{-5/6}P^{-59/12}L_{\rm irr}^{33/16}.   
\end{equation}

The magnetic braking time-scale is given by combining equations \eqref{eq:t_mag} and \eqref{eq:B}:
\begin{equation}\label{eq:t_mag_anal}
\begin{split}
t_{\rm mag}&\propto m^{10/27}P^{-2/9}{\tilde L}_{\rm MeV}^{-4/9}L_{\rm int}^{-4/9}\\
&\propto m^{-20/27}P^{-19/9}{\tilde L}_{\rm MeV}^{-4/9}L_{\rm irr}^{11/12},
\end{split}
\end{equation}
where we have again assumed Roche-lobe overflow, and substituted $L_{\rm int}(L_{\rm irr})$ from equation \eqref{eq:L_int_anal}. We also define
\begin{equation}\label{eq:Lgamma_tilda}
{\tilde L}_{\rm MeV}\equiv L_{\rm MeV}\beta^3,   
\end{equation}
indicating that $L_{\rm MeV}$ is highly degenerate with the magnetic field normalization (magnetic braking by the $L_{\rm MeV}$-driven ablated wind is only sensitive to the combination $L_{\rm MeV} B^3$). 

\begin{figure}
\includegraphics[width=\columnwidth]{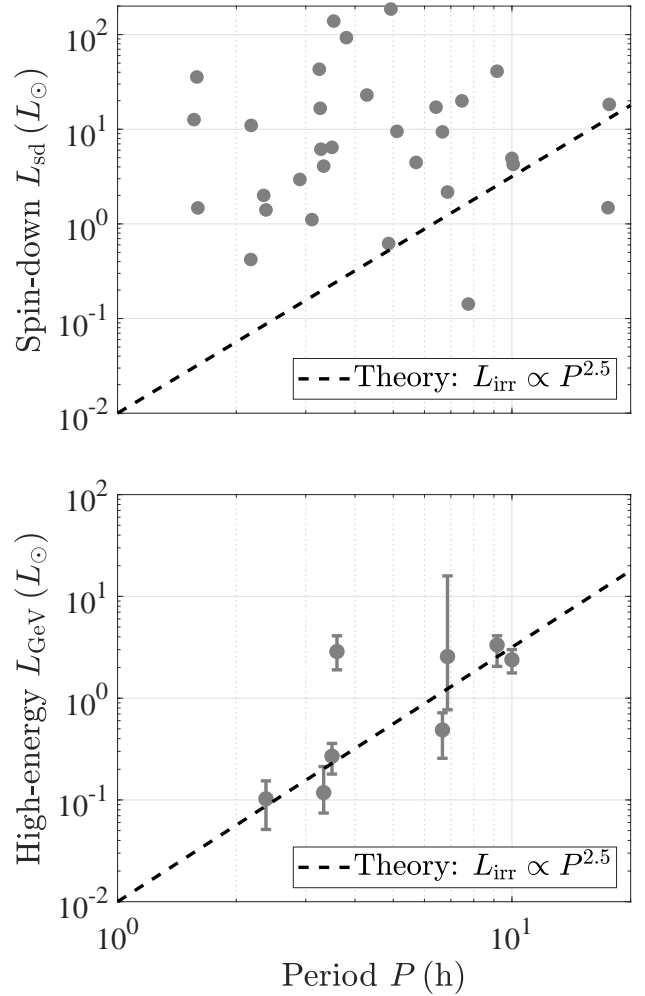}
\caption{In our model, the pulsar's irradiation luminosity is correlated with the orbital period $L_{\rm irr}\propto P^{2.5}$ (dashed black line, given by equation \ref{eq:L_irr_P}), which is calibrated by the evolutionary tracks in Fig. \ref{fig:tracks}. We compare this relation to the measured pulsar luminosities. {\it Top panel}: the pulsar's total spin-down power $L_{\rm sd}$ generally exceeds the minimal required energy (exceptions are discussed in the text). {\it Bottom panel}: the pulsar's high-energy 0.1--100 GeV luminosity \citep[$L_\gamma$ in][denoted by $L_{\rm GeV}$ here, not to be confused with $L_{\rm MeV}$]{Abdo2013} generally agrees with the predicted $L_{\rm irr}(P)$.}
\label{fig:budget}
\end{figure}

The $P(m)$ tracks are given by comparing equations \eqref{eq:t_cool_anal} and \eqref{eq:t_mag_anal}, i.e. demanding that $t_{\rm KH}\sim t_{\rm mag}$: 
\begin{equation}\label{eq:track_anal}
P\propto m^{-10/303}{\tilde L}_{\rm MeV}^{16/101}L_{\rm irr}^{165/404}\simeq m^{-0.03}{\tilde L}_{\rm MeV}^{0.16}L_{\rm irr}^{0.41},
\end{equation}
and the evolution time along the tracks (i.e. age $t$) is given by plugging equation \eqref{eq:track_anal} into either equation \eqref{eq:t_cool_anal} or \eqref{eq:t_mag_anal}
\begin{equation}\label{eq:time_anal}
\begin{split}
t\sim t_{\rm KH}\sim t_{\rm mag}\propto m^{-610/909}{\tilde L}_{\rm MeV}^{-236/303}L_{\rm irr}^{11/202}\\ \simeq m^{-0.67}{\tilde L}_{\rm MeV}^{-0.78}L_{\rm irr}^{0.05}.
\end{split}
\end{equation}
Though less accurate, equations \eqref{eq:track_anal} and \eqref{eq:time_anal} qualitatively reproduce many of the trends we found numerically in Figs \ref{fig:tracks} and \ref{fig:Lgamma}. Equation \eqref{eq:track_anal} shows that $P(m)$ tracks are roughly horizontal, with a period $P\propto L_{\rm irr}^{0.41}$, which is almost insensitive to $L_{\rm MeV}$. The evolution time as given by equation \eqref{eq:time_anal}, on the other hand, is governed by $L_{\rm MeV}$ and does not depend on $L_{\rm irr}$. The analytical equations thus explain the orthogonality of our free parameters --- $L_{\rm irr}$ sets the orbital period, whereas $L_{\rm MeV}$ (or more precisely, $L_{\rm MeV}\beta^3$) sets the age. Equivalently, equations \eqref{eq:track_anal} and \eqref{eq:time_anal} show that the time that a black widow spends at a given mass is $\propto m^{-2/3}$ and nearly independent of its orbital period. This can explain the observational result (Fig. \ref{fig:mdot}) that the lower mass limit to observed black widows is not a strong function of orbital period. Equations \eqref{eq:L_int_anal} and \eqref{eq:track_anal} show that $L_{\rm int}\propto L_{\rm irr}^{-0.33}$, consistent with the weak dependence on $L_{\rm irr}$ in Fig. \ref{fig:Lint} for low $m$ --- stronger irradiation is offset by longer periods. Similarly, $T\propto L_{\rm irr}^{1/4}P^{-1/3}\propto L_{\rm irr}^{0.11}$, explaining why our simulated black widow companions have similar effective temperatures.

\subsection{Energy budget}\label{sec:budget}

For black widows (i.e. low masses), our model predicts a correlation between the pulsar's irradiation luminosity (that is deposited deep in the companion's atmosphere) and the orbital period. We normalize the analytical equation \eqref{eq:track_anal} using our numerical results (Fig. \ref{fig:tracks}): 
\begin{equation}\label{eq:L_irr_P}
L_{\rm irr}\approx 10^{-2}{\rm L}_{\sun}\left(\frac{P}{{\rm h}}\right)^{2.5}.    
\end{equation}
We compare this relation to the observed black widow sample ($m< 10^{-1}{\rm M}_{\sun}$) in Fig. \ref{fig:budget}. The pulsar's spin-down power (taken from the ATNF catalogue), which sets the total energy budget, exceeds the required $L_{\rm irr}$, with two exceptions: PSR J2322$-$2650 is an extremely low-mass outlier, with a minimum mass below $10^{-3}{\rm M}_{\sun}$; PSR J1745+1017 has an atypically long period of 18 h. The latter could survive for Gyrs at such a long period if it under-fills its Roche lobe, which could be tested observationally \citep{Draghis2019}.

Despite providing sufficient energy to power the irradiation, the pulsar's spin-down power does not seem to correlate with the theoretically inferred $L_{\rm irr}$ (top panel of Fig. \ref{fig:budget}). A better agreement is found when comparing directly to the high-energy photon luminosity that was measured by {\it Fermi} for a subset of black widow pulsars (bottom panel). Millisecond pulsars exhibit a range of efficiencies in converting their spin-down power into high-energy photons \citep[fig. 10 in][]{Abdo2013} that can deposit their energy deep enough in the companion's atmosphere to affect its thermal evolution. The outlier J1810+1744 in the bottom panel of Fig. \ref{fig:budget}, which is significantly over-luminous compared to our predicted $L_{\rm irr}(P)$ relation, could perhaps indicate inefficient day--night energy transport; this could be tested by placing tighter constraints on the companion's night-side temperature \citep{Breton2013}. Overall, the agreement in Fig. \ref{fig:budget} between $L_{\rm irr}$, as inferred from evolutionary models, and the measured high-energy luminosities is encouraging, and provides an additional independent validation of our theory.

\section{Conclusions}\label{sec:conclusions}

Black widow companions are thought to be the remnants of main sequence stars that lose most of their mass through a combination of Roche-lobe overflow, driven by magnetic braking, and ablation by the host pulsar's irradiation \citep{Benvenuto2012,Benvenuto2015MNRAS,Chen2013}. While we agree with this general picture, we argue that the roles of the two processes --- magnetic braking and ablation --- are very different from previous studies.
We found in \citetalias{PaperI} that ablation (evaporation) by the pulsar's $\gamma$-ray radiation is too weak to cause significant mass loss on it own, in contrast to previous studies that considered evaporative winds with a free efficiency parameter. In particular, the fraction of the pulsar's spin-down power that goes into driving a wind from the companion is much larger in previous works \citep{Stevens1992, Benvenuto2012} than in our hydrodynamic calculation in \citetalias{PaperI}. When it couples to the companion's magnetic field, however, the ablated wind carries away angular momentum and maintains stable Roche-lobe overflow. We suggest that this ablation-driven magnetic braking is more appropriate for understanding black widows than previous prescriptions that extrapolated empirical magnetic braking relations fit for isolated (non-irradiated) stars. 

In this paper we used \textsc{mesa} to compute consistent evolutionary tracks, in which an initially main sequence solar mass star is reduced, over several Gyr, to a few per cent of its original mass. Our model assumes by construction stable Roche-lobe overflow and has two free parameters: the MeV $\gamma$-ray pulsar luminosity $L_{\rm MeV}$ which drives the evaporation, and the broader spectrum $L_{\rm irr}\gtrsim L_{\rm MeV}$ that deposits energy deeper in the companion's atmosphere. We find that this deeper energy deposition is necessary to explain long-lived black widows that have large radii and fill (or almost fill) their Roche lobes. By changing the companion's outer boundary condition, $L_{\rm irr}$ reduces its internal luminosity $L_{\rm int}$, similar to irradiated hot Jupiters \citep{Guillot1996,ArrasBildsten2006}. The slower cooling allows black widow companions to remain inflated, and thus fill their Roche lobes at longer orbital periods $P$. Different values of $L_{\rm irr}$ can therefore reproduce the period range spanned by observed black widow systems (Fig. \ref{fig:tracks}).  

We find that the evolutionary tracks of black widows are determined by a balance between the companion's Kelvin--Helmholtz cooling time $t_{\rm KH}$ and the mass loss time-scale $m/\dot{m}$.
In stable Roche-lobe overflow, mass is lost at the same rate as angular momentum, which is removed by magnetic braking and, at short orbital periods, gravitational waves. Magnetic braking requires two ingredients: a mass outflow (wind), which we calculated in \citetalias{PaperI}, and a magnetic field $B$. Previous studies \citep{Chen2013, DeVito2020} assumed that magnetic fields, and as a consequence, magnetic braking, cease to operate when companion stars lose their radiative cores, at about $0.3 {\rm M}_{\sun}$. Observations and theory, however, indicate that even low-mass fully convective objects can generate strong fields, powered by the convective luminosity $L_{\rm int}$. Here, we assumed that $B(L_{\rm int})$ is given by the \citet{Christensen2009} relation, which successfully reproduces many of these observations. Moreover, we find that due to their lower internal heat transport $L_{\rm int}$, strongly irradiated companions retain radiative cores down to black widow masses $m\sim 10^{-2}{\rm M}_{\sun}$ (Fig. \ref{fig:convective}) --- magnetic braking thus likely persists for the entire evolution of such systems. Our results are also not sensitive to a modest decrease in $B$ when the companion becomes fully convective (Fig. \ref{fig:convective}). Such a model may in fact be favoured by the excess of black widows relative to redbacks.

The observed black widow period and mass distributions are fit well by $L_{\rm MeV}= 0.05-0.1\, {\rm L}_{\sun}$ and $L_{\rm irr}= 0.1-3\, {\rm L}_{\sun}$. The $L_{\rm irr}$ range agrees with {\it Fermi} measurements \citep{Abdo2013}, which are consistent with the approximate $L_{\rm irr}\propto P^{2.5}$ correlation that we find both analytically and numerically (Fig. \ref{fig:budget}). The large $L_{\rm MeV}/L_{\rm irr}$ range indicates that millisecond pulsars might have diverse spectra. As an independent test, we compared the night-side temperatures of our simulated black widow companions with optical light curves \citep{Draghis2019}. In our models, these temperatures are set directly by $L_{\rm irr}$ and span a narrow range around $3000\textrm{ K}$, consistent with most well-constrained observations (Fig. \ref{fig:Draghis}). The optical light curves also indicate that most black widow companions fill or almost fill their Roche lobes, in agreement with the sustained magnetic braking scenario.

Our black widow evolutionary tracks also form more massive redback companions at earlier times (Fig. \ref{fig:tracks}). This is in contrast to some previous studies \citep{Chen2013}, where the two spider populations evolve primarily on separate tracks (see, however, \citealt{Benvenuto2014} and some of the tracks in \citealt{DeVito2020}). While we reproduce the short-period redback population, we find it difficult to explain redbacks with orbital periods longer than about 8 h; these require $L_{\rm irr}>10{\rm L}_{\sun}$ --- about 100 per cent of 
the host pulsar's spin-down power $L_{\rm sd}$ (all redbacks with $P>8\textrm{ h}$ have $L_{\rm sd}>10{\rm L}_{\sun}$). Another possible energy source is (past or present) accretion from the donor onto the pulsar. Our simplest models also do not reproduce the tentative mass gap between black widows and redbacks \citep{Roberts2013, Chen2013}. This might be related to the gap in stellar rotation periods \citep{Meibom2011} --- both gaps can be explained by a bimodal magnetic field distribution. Specifically, solitary stars with strong magnetic fields spin down through magnetic braking to long rotational periods, whereas weakly magnetized stars remain fast rotators. Similarly, pulsar companions with strong fields evolve to become black widows, while weakly magnetized companions remain redbacks.    

On the opposite side of the companion mass spectrum, the two extremely low-mass companions to PSR J1719$-$1438 \citep{Bailes2011} and to PSR J2322$-$2650 \citep{Spiewak2018} also challenge our model. According to equations \eqref{eq:Lgamma_tilda} and \eqref{eq:time_anal}, main sequence stars can be reduced to such low masses ($m\sim 10^{-3}{\rm M}_{\sun}$) in several Gyr if their magnetic fields are stronger than our nominal $\beta=1$ by a factor of a few (confirmed by \textsc{mesa}). Alternatively, these companions might have formed in a bottom-up scenario, perhaps similarly to the much smaller `pulsar planets' around PSR B1257+12 \citep{Lin1991,PhinneyHansen93,Menou2001,MargalitMetzger2017}.

\section*{Acknowledgements}

We thank Jim Fuller for suggesting the relation between the spider mass gap and the stellar rotation period gap. SG thanks the astrophysics group at the Weizmann Institute of Science for stimulating discussions.
We thank the \textsc{mesa} team for this valuable tool, and specifically Evan Bauer for making it easily accessible to \textsc{windows} users through \textsc{mesa-docker}.
We thank Lars Bildsten, Jim Fuller, Amir Levinson, Roger Romani, and the anonymous reviewer for comments on the draft which improved the paper.
SG is supported by the Heising-Simons Foundation through a 51 Pegasi b Fellowship. This research benefited from interactions at ZTF Theory Network meetings, funded by the Gordon and Betty Moore Foundation through Grant GBMF5076. 

\section*{Data availability}
The data underlying this article will be shared on reasonable request to the corresponding author.




\bibliographystyle{mnras}
\input{bw.bbl}




\bsp	
\label{lastpage}
\end{document}